\documentstyle[preprint,aps,amsfonts]{revtex}
\def \A {{\cal A}} 
\def \F {{\cal F}}
\begin{document}
\draft
\title{ \hfill{\normalsize  hep-th/9610046
} \\[-14pt]\hfill {\normalsize CALT-68-2074}\\[-16pt]
\hfill {\normalsize QMW-PH-96-23}\\[-16pt]
\hfill {\normalsize UCSBTH-96-21}\\[-16pt]
\hfill{\normalsize NSF-ITP-96-77 } \\
\smallskip
Entropy of Four-Dimensional Rotating
BPS Dyons}
\author{Jerome P. Gauntlett$^{a}$ and Tsukasa Tada$^{b}$
\footnote{Present address: KEK Theory Group, 1-1 Oho Tsukuba Ibaraki
305, Japan.}}
\address{\  \\
 ${}^a$ California Institute of Technology\\
Lauritsen Laboratory, Pasadena, CA 91125\\
and \\
Department of Physics \\
 Queen Mary and Westfield College \\
Mile End Road, London E1 4NS, U.K.\\
 j.p.gauntlett@qmw.ac.uk\\
\  \\
${}^{b}$ ITP and Department of Physics \\
University of California \\
Santa Barbara, CA 93106, USA \\
tada@theory.kek.jp
 }

\maketitle
\vfill\eject
\vskip 2cm

\hskip 3cm
\begin{abstract}
The known BPS dyon black hole solutions of the $N=4$ heterotic
string in four dimensions with a non-zero angular momentum
all have naked singularities. We show that it is possible to modify
a certain class of these solutions by the addition of massive Kaluza-Klein
fields in such a way that the solutions decompactify near the core to
five-dimensional black hole solutions with regular event horizons.
We argue that the degeneracy of the four-dimensional BPS dyon states
is given, for large charges, by the five-dimensional geometric entropy.
\end{abstract}

\vfill \eject

There has recently been substantial progress in understanding the relationship
between string states and classical black hole solutions. (For example, see
\cite{1,2}
and references therein.)
In this Rapid Communication we would like to address some further issues involving
spin within the
context of the $N=4$ heterotic string compactified on a six-torus.
In particular, we will argue that the degeneracy of a certain
class of Bogomol'nyi-Prasad-Sommerfield (BPS) dyon states with a non-zero angular momentum can be
determined from classical solutions.

It was shown
in \cite{dghw,cmp} that perturbative electrically charged BPS states can be
identified with classical solutions by allowing massive Kaluza-Klein
modes to be excited. At large distances the solutions
appear to be
four-dimensional rotating electrically charged BPS black holes
but at
shorter distances they decompactify to five-dimensional string solutions 
wrapped around a circle that have 
traveling waves of arbitrary profile. If the massive Kaluza-Klein fields
are ignored the solutions have naked singularities. When they are included,
the solutions still have naked singularities but it was shown in \cite{dghw}
that they have a natural interpretation
in that they map
onto an appropriate string source.
{}From this point of view the degeneracy of BPS string states
simply corresponds to the different wave profiles that the string solutions
can possess.

The perturbative BPS states all preserve one-half of the supersymmetry. The
conjectured $S$ duality of the four-dimensional heterotic theory
maps these states onto dyonic states whose electric and magnetic charge
vectors are parallel. $S$ duality predicts that all of the dyonic BPS states
on the orbit of a given perturbative state have
the same degeneracy. Some partial checks of this have been made
in \cite{ss,porrati}. 

General dyonic BPS states whose electric and magnetic charge vectors are not
parallel preserve one quarter of the supersymmetry.
Black hole solutions with such charges based on four independent harmonic
functions have been constructed
which have event horizons with non-zero area \cite{cvetic}. It is
natural to
interpret the geometric entropy of these black holes as
being the degeneracy of dyonic states
in the string Hilbert space with the same charges,
in the limit that the charges are large. Indeed the duality between 
the heterotic string on $T^6$ and the type IIA string on $K3\times T^2$ 
and $D$-brane state counting in the type IIA
theory support this interpretation \cite{ms,jkm}.
At present it is not clear how to
extend this picture to include dyonic BPS states with non-zero angular
momentum as the corresponding classical solutions all
exhibit naked singularities
\cite{cveticr}.

The dyonic BPS solutions based on three harmonic functions also
preserve one quarter of the supersymmetry but they have naked singularities
(these are just the solutions of \cite{cveticr} with one of the four harmonic
functions set to zero).
We will argue that these classical
solutions can be dressed with massive Kaluza-Klein fields so that at shorter
distances they decompactify to five-dimensional black holes with regular
event horizons. We would like to identify the five-dimensional geometric 
entropy as
corresponding, for large charges,
to the degeneracy of the corresponding four-dimensional dyonic
states. This procedure can also be extended to include rotation: after the
inclusion of the massive Kaluza-Klein modes the solutions decompactify
to five-dimensional black holes with $J^{12}=J^{34}$, where $J^{12}$ and $J^{34}$  
are
the angular momentum in two orthogonal planes.

The procedure for adding the massive Kaluza-Klein massive modes is
straightforward. We begin with
five-dimensional BPS black holes
with $J^{12}=J^{34}$ which have regular event horizons \cite{bmpv,2}.
By constructing a periodic array of these black holes, we construct
four-dimensional black holes that have non-trivial dependence
on the compactified
coordinate.
For large radii in four dimensions
the solutions appear as four-dimensional rotating dyonic black holes based
on three independent harmonic functions. At small distances the solutions
decompactify into five-dimensional black holes.

The idea of determining the entropy of
singular four-dimensional black holes by adding massive Kaluza-Klein
modes was also considered in \cite{bb}.
There it was shown that dilatonic black holes can be uplifted
to regular solutions in higher dimensions.
(Related ideas were explored in 
\cite{ght,hs}). For dilaton coupling parameter $a$ given by
$a=0,1,{\sqrt 3}$ it was shown that an entropy formula with the correct
scaling behaviour could be assigned to the corresponding
four-dimensional black holes
by calculating the area of an $S^2$ that intersects the regular horizon
in higher dimensions.
However, their procedure
did not work for the case $a=1 /\sqrt 3$, which is the
case most similar to ours, and moreover they did not consider rotation.

Our starting point is the generalized six-dimensional chiral null model
\cite{ht,ct,chiral6d} :
\begin{eqnarray}
L&=&F(x)\partial u \left[ \bar\partial v + K(x) \bar\partial u+ 2 \A_m(x)
\bar\partial x^m \right]
\nonumber \\
&&+(g_{mn}+B_{mn})(x)\partial x^m \bar\partial x^n +\partial y_a \bar\partial
y_a+{\cal R}(\frac12 \ln f + \frac12 \ln F) ,\label{6dchiralmodel}
\end{eqnarray}
where $ u=y_5-t, v=y_5+t$, $\ \{x^m\} \ \ (m=1,2,3,4)$ are non-compact
spatial coordinates
and $ \{y_5,y_a\} \ (a=6,7,8,9)$ are the coordinates
of a 5-torus
\footnote{Note that in the notation of \cite{chiral6d}
we have set $\A_a =0$ and dropped the dependence
of $K$ and $\A_m$ on $u$ as this is sufficient for our purposes.}.
In addition we have
\begin{equation}\label{eqgH}
g_{mn}=f(x)\delta_{mn},\  \ H_{mnk}=-\epsilon_{mnkp}\partial_p f,
\end{equation}
where $ H_{mnk} \equiv 3 \partial_{[m}B_{nk]}$ and $\epsilon_{mnkp}$ stands
for flat space anti-symmetric tensor.
The conditions for exact conformal invariance are given by
$$
\partial^m \partial_m f=0, \ \ \partial^m \partial_m F^{-1}=0,
\ \ \partial^m \partial_m K =0,
$$
\begin{equation}\partial^m (f^{-1}\F_{+mn})=0,
\label{eqfFKF+}
\end{equation}
where $ \F_{+mn}$ is the self dual part of
the field strength $ \F_{mn} \equiv \partial_m \A_n -\partial_n \A_m$.

A five-dimensional rotating BPS black hole solution is obtained by taking
\begin{eqnarray}
&&f=1+\frac{P}{\rho ^2}, \ F^{-1}=1+\frac{Q_2}{\rho^2}, \ K=1+\frac{Q_1}{\rho
^2}, \nonumber \\
&& \label{5dsingle} \\
&& \A_m(x) dx^m = \frac{\gamma}{\rho ^4} (-x^2
dx^1+x^1dx^2-x^4dx^3+x^3dx^4), \nonumber
\end{eqnarray}
where $ \rho^2 \equiv x^mx^m$ \cite{2}.
Dimensional reduction along the $y_5$ direction leads to the five-dimensional
Einstein metric:
\begin{eqnarray}
ds^2_{E5}&=&-\lambda^2(\rho) (dt+\A_m dx^m)^2+\lambda^{-1}(\rho) dx^m dx_m,
\nonumber \\
\lambda(\rho)&\equiv&(fF^{-1}K)^{-1/3},
\label{metric}
\end{eqnarray}
along with $U(1)$ gauge fields $A^{(1)}_{ 5 \mu}$
(Kaluza-Klein gauge field from $y_5$ ) and $A^{(2)}_{5 \mu }$
(gauge field arising from $y_5$ component of the ten-dimensional
two-form field) whose electric charge is $-Q_1$ and $-Q_2$ respectively.
There is also a charge defined by
$\frac{1}{4\pi^2} \int_{S^3}{H}$ whose value is $-P$.
The quantization of these charges can be determined from
\cite{dghw,rw}:
\begin{equation}
Q_1=\frac{4 G^{(5)}_N}{\pi R_5}m , \ \ 
Q_2=\frac{4 G^{(5)}_N}{\pi\alpha '}wR_5, \ \
P=\alpha ' n.\\ \label{Q1Q2qnt}
\end{equation}
Here $R_5$ is the radius of the $y_5$ circle and $w,m,n\in {\Bbb Z}$.
Also we denote the $D$-dimensional Newton constant by $G_N^{(D)}$.

Using the formulae in \cite{mp} the ADM mass and the angular
momenta of the black hole can be determined from the asymptotic
form of the metric:
\begin{eqnarray}
M_5&=&\frac{\pi}{4G_N^{(5)}} (Q_1+Q_2+P),\  \ \nonumber\\
&& \label{MJ5} \\
J^{12}&=&J^{34}=\frac{\pi \gamma}{4G_N^{(5)}} \equiv J. \nonumber
\end{eqnarray}
The charges and the mass formula imply that the solution saturates a BPS  
bound
and preserves one quarter
of the supersymmetry \cite{cvetic5,dlr}.
The horizon of the black hole is located at $ \rho=0$ and
has non-zero horizon area $A_5$. The
corresponding Bekenstein-Hawking entropy is given by
\begin{eqnarray}\label{entropy}
S=\frac{1}{4G_N^{(5)}} A_5 &=&
\frac{2\pi^2}{4G_N^{(5)}} [Q_1Q_2P-\gamma^2]^{1/2} \nonumber \\
&=&2\pi[ nmw -J^2]^{1/2}.
\end{eqnarray}
Note that the entropy is expressed in terms of integers
which are independent of moduli. Requiring that the horizon remains
euclidean imposes the bound $J^2\le nmw$ on the angular momentum.

We now turn to the construction of four-dimensional solutions
that are dressed with massive Kaluza-Klein
fields. Since the five-dimensional black holes are BPS saturated
we can construct a periodic array along, for instance, the $x^4$ direction
to obtain:
\begin{eqnarray}
f&=&1+\sum_{k=-\infty}^{k=\infty}\frac{P}{r^2+(x^4-2\pi R_4k)^2}
=1+\frac{P}{2R_4r}\sinh \frac{r}{R_4} \left( \cosh \frac{r}{R_4} -\cos
\frac{x^4}{R_4}  \right)^{-1}, \nonumber \\
F^{-1}&=&1+\sum_{k=-\infty}^{k=\infty}\frac{Q_2}{r^2+(x^4-2\pi R_4k)^2}
=1+\frac{Q_2}{2R_4r}\sinh \frac{r}{R_4} \left( \cosh \frac{r}{R_4} -\cos
\frac{x^4}{R_4}  \right)^{-1}, \nonumber \\
K&=&1+\sum_{k=-\infty}^{k=\infty}\frac{Q_1}{r^2+(x^4-2\pi R_4k)^2}
=1+\frac{Q_1}{2R_4r}\sinh \frac{r}{R_4} \left( \cosh \frac{r}{R_4} -\cos
\frac{x^4}{R_4}  \right)^{-1},\nonumber \\
\A_m dx^m& =&\sum_{k=-\infty}^{k=\infty}\frac{\gamma}{\left( r^2+(x^4-2\pi
R_4k)^2 \right)^2}\left(-x^2
dx^1+x^1dx^2+x^3dx^4\right),
\label{eqfFKAmulti}
\end{eqnarray}
where $ r^2\equiv (x^1)^2+(x^2)^2+(x^3)^2$.
The solution is now periodic in the $ x^4$
direction and so we can take the $ x^4$ direction to be compactified
with radius $R_4$.
The four-dimensional Einstein metric and various
fields can be determined using the standard procedure \cite{mahsch}.
Defining
\begin{equation}
{\bar \lambda} \equiv \frac{F}{\Delta^{1/2}}, \ \  \Delta \equiv fFK -\A_4^2
F^2,
\end{equation}
we obtain
\begin{eqnarray}
&&
ds_{E4}^2=-{\bar \lambda} [dt + \A_i dx^i ]^2 +{\bar \lambda}^{-1} (dx^i)^2
\ \ \ \ \ (i=1,2,3), \nonumber \\
&&A_{4t}^{(1)}=-\frac{\A_4F^2}{\Delta}, \ \ \ A_{4i}^{(1)}=-\frac{\A_4F^2
}{\Delta}\A_i, \nonumber \\
&&
A_{5t}^{(1)}=-1+\frac{fF}{\Delta}, \ \ \ A_{5i}^{(1)}=\frac{fF }{\Delta}\A_i,
\label{mess}
\\
&&
A_{4t}^{(2)}=\frac{\A_4fF^2}{\Delta}, \ \ \ A_{4i}^{(2)}=B_{4i}+\frac{\A_4
fF^2} {\Delta}\A_i,
\nonumber \\
&& A_{5t}^{(2)}=\frac{fF^2K}{\Delta}, \ \ \ A_{5i}^{(2)}=\frac{fF^2K
}{\Delta}\A_i,
\nonumber \\
&&H'_{ijk}=\frac{\A_4F^2}{\Delta}\left( \A_i\epsilon_{jkl}\partial_l f +
(\hbox{cyclic permutations on $i,j,k$}) \right), \nonumber \\
&&e^{2\phi}={fF\over \Delta^{1/2}}, \nonumber 
\end{eqnarray}
where $\A^{(1)}_{4,5}$ are the Kaluza-Klein gauge fields coming coming from the
$x^4,y^5$ circles, $\A^{(2)}_{4,5}$ are the gauge fields arising from
the $x^4,y^5$ components of the ten-dimensional two-form,
$H'$ is the gauge invariant three-form in four dimensions \cite{mahsch}, $\phi$
is the four dimensional dilaton, 
$\partial_i B_{4j}-\partial_j B_{4i}=-\epsilon_{ijk}\partial_k f$
and $
\epsilon_{ijk}$
is flat space anti-symmetric tensor.
The asymptotic form of the harmonic functions and $ \A_m$ are
given by
\begin{eqnarray}
&& f\approx 1+\frac{\bar P}{r}, \ \ F^{-1}\approx 1+\frac{\bar Q_2}{r}, \ \
K\approx 1+\frac{\bar Q_1}{r} \ \ \nonumber
\\ &&  \label{fFKA4d}\\
&&\A_m dx^m \approx
\frac{{\bar\gamma}}{r^3}(-x^2dx^1+x^1dx^2+x^3dx^4),\nonumber
\end{eqnarray}
where we have defined
\begin{equation}
{\bar P}\equiv \frac{P}{2R_4}, \ {\bar Q_1}\equiv \frac{Q_1}{2R_4} ,
\ {\bar Q_2}\equiv \frac{Q_2}{2R_4}, \ \ {\bar \gamma}\equiv
\frac{\gamma}{4R_4}.
 \label{eqbar}
\end{equation}

The ADM mass and the angular momentum of this four-dimensional
solution are exactly the same as
the five-dimensional black hole
and can be expressed in terms of four-dimensional quantities as:
\begin{eqnarray}
&&M_{ADM}=\frac{1}{4G_N^{(4)}} ({\bar P}+{\bar Q_1}+{\bar Q_2}),
\nonumber \\
\label{4dMJ}\\
&&J^{12}=\frac{{\bar \gamma}}{2G_N^{(4)}}. \nonumber
\end{eqnarray}
In addition the charges and dipole moments of the gauge fields are given by
\begin{eqnarray}
&&A^{(1)}_4  \ \ \  \hbox{electric dipole moment}
\ \ \   {\bar \gamma}; \nonumber \\
&&A^{(1)}_5  \ \ \ \hbox{electric charge} \  -{\bar Q_1} , \
\hbox{magnetic dipole moment } \ \
-{\bar \gamma}; \nonumber \\
&&A^{(2)}_4  \ \ \ \hbox{magnetic charge}  \ \ {\bar P} ,
\ \ \hbox{electric dipole moment} \ \
-{\bar \gamma}; \label{4dcharges}\\
&&A^{(2)}_5  \ \ \ \hbox{electric charge}\   -{\bar Q_2} ,
\ \ \hbox{magnetic dipole moment} \ \
-{\bar \gamma}. \nonumber
\end{eqnarray}
Note that from a four-dimensional point of view
the charge quantization conditions are given by \cite{dghw,rw}
\begin{equation}
{\bar Q_1}=\frac{4 G^{(4)}_N}{R_5}m, \ \
{\bar Q_2}=\frac{4 G^{(4)}_N}{\alpha '}wR_5, \ \
{\bar P}={\alpha'\over 2 R_4} n,
\end{equation}
which, after using Eq. (\ref{eqbar}), coincides with Eq. (\ref{Q1Q2qnt})
as they should.

The asymptotic form of our four-dimensional solution
given by Eqs. (\ref{mess}) and (\ref{fFKA4d}) is in fact a solution
for all values of $r$ and was
constructed by Cveti\v{c} and Youm in
\cite{cveticr} (i.e., by setting one of the four harmonic functions in
the solution in \cite{cveticr} to zero).
This BPS solution has a naked singularity and
consequently it is not clear how the entropy of states
with charges as in (\ref{4dcharges}) can be determined from it.
However, the solution (\ref{mess}),(\ref{eqfFKAmulti}), which
asymptotically
approaches the Cveti\v{c} and Youm solution but is dressed with massive
Kaluza-Klein fields in such a way that it decompactifies to a five-dimensional
black hole, allows us to address this issue. Our proposal is
that the degeneracy of states,
for large values
of the charges, is simply given by
the five-dimensional geometric entropy  (\ref{entropy})
\footnote{This also agrees with the suggestion presented in \cite{dvv}.}.

We now conclude with some additional comments.
Cveti\v{c} and Youm \cite{cvetic5} have shown that the five-dimensional
black hole solution based 
on three harmonic functions can be used to generate all
five-dimensional black holes with regular event horizons by
using $O(5,21)$ transformations. Consequently the procedure
we have described to add in massive Kaluza-Klein modes to
the four-dimensional singular black holes can be generalized to
all four-dimensional black holes that can be obtained using $O(5,21)$
transformations.

String-string duality in six-dimensions leads to the heterotic string
compactified on $T^6$ being dual to the type II string compactified
on $K3\times T^2$. How are our results interpreted from the type II point
of view? The four-dimensional singular black hole based on
three harmonic functions can be interpreted
as a D-6-brane intersecting a D-2-brane in a membrane with
momentum flowing along one of the membrane directions.
At weak coupling the number of D-brane states with the 
appropriate charges can be determined using string perturbation theory.
Acting with T-duality we obtain a D-5-brane intersecting a 
D-1-brane with open strings carrying momentum along the string
intersection. But this is precisely the situation that leads
to a degeneracy of states that gives the entropy of the corresponding
five-dimensional black holes \cite{sv,cm,bmpv}.
Thus, it is consistent that the  
singular four-dimensional black hole solution can be modified by
including massive Kaluza-Klein modes to decompactify to a
five-dimensional black hole near the core, as we have shown. 

It is known that in certain limits the 
entropy of non-extremal five-dimensional
black holes can also be given a state counting interpretation 
using D-branes \cite{cm,HS,blmpsv}. The above correspondence suggests that it might
be possible to add massive Kaluza-Klein modes to non-extremal
four-dimensional black holes with three charges,
in such a way that they decompactify
onto five-dimensional non-extreme black holes. Clearly the 
construction used in this paper cannot be straightforwardly generalized 
as we relied
on the BPS property of the five-dimensional black holes in
order to construct the periodic array. It would be interesting
to see if
more general methods are applicable and some positive indications
of this were recently presented in \cite{lupopexu}.

{\bf Acknowledgments}: We would like to thank 
F. Dowker, G. Horowitz, D. Kastor and D. Lowe for
helpful discussions.
JPG was supported in part by
the U.S. Dept. of Energy under Grant No. DE-FG03-92-ER40701, and
TT was partially
supported by NSF Grant
PHY89-04035 and the JSPS.

\end{document}